\documentclass[final,3p,times,twocolumn]{elsarticle}
\usepackage{amssymb, lineno, hyperref, subcaption}
\journal{Solid State Communications}

\begin{document}

\begin{frontmatter}

\title{Magnetic phase diagram of TbMn$_{1-x}$Fe$_x$O$_3$ $(0 \leq x \leq 1)$ substitutional solid solution}

\author[addr1]{M. Mihalik jr.\corref{cor1}}
\cortext[cor1]{Corresponding author} 
\ead{matmihalik@saske.sk}
\author[addr1]{M. Mihalik}
\author[addr2]{Z. Jagli\v ci\'c}
\author[addr3]{R. Vilarinho}
\author[addr3]{J. Agostinho Moreira}
\author[addr4]{E. Queiros}
\author[addr4]{P. B. Tavares}
\author[addr3]{A. Almeida}
\author[addr1]{M. Zentkov\'a}
\address[addr1]{Institute of Experimental Physics Slovak Academy of Sciences, Watsonova 47, Ko\v sice, Slovak Republic}
\address[addr2]{Institute of Mathematics, Physics and Mechanics and Faculty of Civil and Geodetic Engineering, University of Ljubljana, Slovenia}
\address[addr3]{IFIMUP and IN-Institute of Nanoscience and Nanotechnology, Departamento de F\'isica e Astronomia da Faculdade de Ci$\hat{e}$ncias, Universidade do Porto, Porto, Portugal}
\address[addr4]{Centro de Qu\'imica, Universidade de Tras-os-Montes e Alto Douro, Vila Real, Portugal}

%
%
\begin{abstract}
We present the magnetic phase diagram of TbMn$_{1-x}$Fe$_x$O$_3$ substitutional solid solution in the whole concentration range $0 \leq x \leq 1$ as determined from magnetization and specific heat measurements. We have found that the dominant magnetic ion in the concentration range $0 \leq x < 0.3$ is manganese, while iron ions do not create independent magnetic structure, but strongly affect magnetic properties of the parent compound by reducing transition to magnetically ordered state and transition into a cycloidal phase. The magnetism in the concentration range $0.3 < x \leq 1$ is driven by the Fe sublattice. The manganese ions again do not order in long range magnetic ordered state, but stabilize four different magnetic structures of Fe sublattice above 2 K. The magnetic ordering of Tb sublattice was observed only on parent compounds TbMnO$_3$ and TbFeO$_3$ and for doping level below 0.1, or over 0.9.
\end{abstract}

\begin{keyword}
A. Manganates \sep A. Orthoferrites \sep B. Doping \sep D. Magnetic phase diagram
\end{keyword}

\end{frontmatter}
%
%

\section{Introduction}
Rare-earth- ($RE$) transition-metal ($TM$) oxides show a great variety of intriguing phenomena originating from the coupling between the two spin subsystems \cite{tokura2010}. In the case of TbMnO$_3$ this interplay led to magnetic ordering of Mn sublattice at $T_{\rm N}$ = 41 K into longitudinal spin density wave which propagates along the $a$-axis, change the magnetic structure into a cycloidal phase with a component along the $b$-axis below $T_{\rm S}$ = 28 K and, finally, the ordering of Tb sublattice below 7 K into non-collinear magnetic structure \cite{aliouane2008}. Since the second magnetic phase transition is accompanied with the spontaneous electric polarization \cite{kimura2005}, the TbMnO$_3$ is very extensively studied in last decade. 

In the aim to understand the basic magnetic and electric phenomena in TbMnO$_3$ and/or to optimize some physical properties, many researches  have doped either RE ions on the Tb site \cite{kumar2014}, or $TM$ ion on either Tb, or Mn site \cite{xu2014, lu2014, cuartero2012,  staruch2014, wang2015}. From these trials especially the Fe$^{3+}$ non Jahn-Teller ion doping on the Mn$^{3+}$ Jahn-Teller ion site looks very interesting. Advantage of this doping is good solubility which allows synthesis of TbMn$_{1-x}$Fe$_x$O$_3$ substitutional solid solution for the whole concentration range $0 \leq x \leq 1$. Since Fe$^{3+}$ has one more unpaired electron in compare with Mn$^{3+}$ and TbFeO$_3$ orders antiferromagnetically at $T_{\rm N}$ = 681 K \cite{gukasov1997}, this doping can increase differences in magnetic interaction between the nearest and the next nearest magnetic ions leading to the enhanced spin frustration in material. Spin reorientation magnetic phase transitions for concentrations $x$ = 0.5; 0.75 and 1 \cite{gukasov1997, nhalil2015, kim2011} was also reported on Fe rich materials. All these facts together with our previous preliminary work \cite{mihalik2015} indicate that the magnetic phase diagram of this system is rather complex. In this paper we present our research of magnetic and thermal properties revealing the magnetic phase diagram of the TbMn$_{1-x}$Fe$_x$O$_3$ system in the whole concentration range $0 \leq x \leq 1$.     
%
%
\section{Sample preparation and experimental methods}
Polycrystalline samples for $x$ = 0; 0.02; 0.04; 0.1; 0.3; 0.5; 0.55; 0.7; 0.95 and 1 were prepared by floating zone method in FZ-T-4000 (Crystal Systems Corporation) mirror furnace. As starting materials we used oxides of MnO$_2$, Tb$_4$O$_7$ (both, purity 3N; supplier: Alpha Aesar) and Fe$_2$O$_3$ (purity 2N, supplier: Sigma Aldrich). The starting materials were mixed in a Tb:Mn:Fe stoichiometric ratio as intended for the final compound, cold pressed into rods and sintered at 1100 $\circ$C from 12 to 14 hours in air. The growing of the bulk samples was performed in air atmosphere. 

Ceramics for concentrations $x$ = 0.075; 0.2; 0.4 and 0.6 were prepared by urea sol-gel combustion method. Details of preparation can be found elsewhere \cite{moreira2010}.
The quality of grown polycrystals was checked by X-ray powder diffraction experiment and by the energy dispersion X-ray analysis. All samples were confirmed to be single-phased within the uncertainty of used experimental methods and with intended chemical composition. 

Specific heat was measured by relaxation method on PPMS (Quantum Design) apparatus on the bulk samples directly cut from the grown ingot. Magnetization and AC susceptibility measurements were performed on powdered samples by MPMS and MPMS3 (both from Quantum Design) apparatuses in the temperature range from 2 K to 300 K. Magnetization measurements above room temperature were performed on PPMS (Quantum Design) apparatus with installed VSM oven option. For these experiments bulk samples fixed to the oven holder by Zircar cement have been used.
%
%
\section{Results and discussion}
%
%
\begin{figure}[t]
\begin{center}
\includegraphics[angle=0,width=0.45\textwidth]{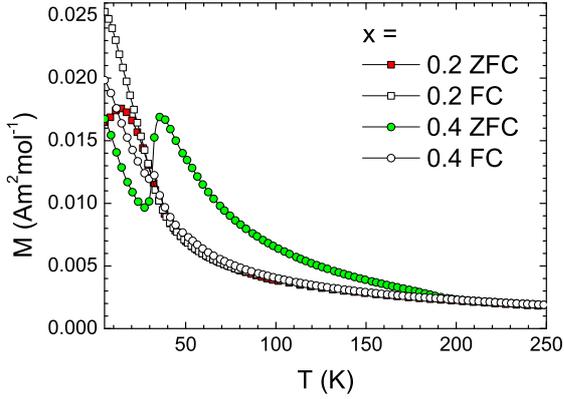}
\caption{The Zero field cooled and field cooled curves for $x$ = 0.2 and $x$ = 0.4 composition. The data were measured with applied magnetic field $\mu_0$H = XXX T. \label{fig1}}
\pdfcompresslevel=9
\end{center}
\end{figure}
%
%

\begin{figure}[t]
\begin{center}
\includegraphics[angle=0,width=0.45\textwidth]{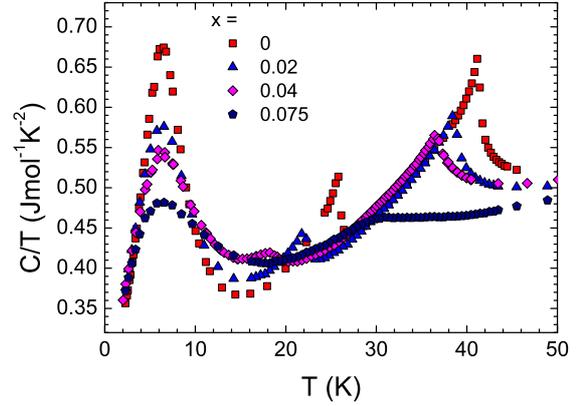}
\caption{The specific heat for low Fe concentrations showing the peaks associated with $T_{\rm N}$ and $T_{\rm S}$ as well as the bump at 7 K which is associated with ordering of Tb sublattice. \label{fig2}}
\pdfcompresslevel=9
\end{center}
\end{figure}

Our previous measurements of magnetization, heat capacity, dielectric permittivity and polar measurements on samples $x$ = 0; 0.01; 0.25; 0.04; 0.05 and 0.1 \cite{mihalik2015, vilarinho2016} revealed that the temperature of $T_{\rm N}$ decreases from 41 K for $x$ = 0 to 32.6(6) K for $x$ = 0.1. Also, $T_{\rm S}$ decreases from 25 K for $x$ = 0 to 17 K for $x$ = 0.04, but was not observed already for $x$ = 0.05 \cite{vilarinho2016}. ZFC-FC curves measured for $x$ = 0.2 (Fig. \ref{fig1}) bifurcate from each other at 30.3 K. This suggests that there exist the paramagnetic-to-magnetic ordering phase transition also for this concentration. We have determined this transition as min($\partial M/\partial T$) = 30.5 K. All previously mentioned anomalies exhibit the same tendency with iron substitution: decrease in temperature with increasing of iron content. That is why we associate them with ordering of Mn sublattice. The opened question remained, what was the concentration evolution of the ordering of Tb sublattice. Since the magnetization experiment does not trace this transition very well, we have decided to determine it from specific heat measurements. The specific heat measurements were partly published in our previous article \cite{vilarinho2016} and the rest of the data are presented in Fig. \ref{fig2}. Aliouane et al. \cite{aliouane2008} reported the ordering of Tb sublattice in TbMnO$_3$ at temperature 7 K. At this temperature range we have observed the bump on $C/T$ vs. $T$ curves and we assign this feature to ordering of Tb sublattice. This feature broadens and smears out with iron doping for $x \leq$ 0.075, but does not shift in temperature. 

%
%
\begin{figure*}[t]
\begin{center}
\begin{subfigure}{0.45\textwidth}
\includegraphics[angle=0,width=0.9\textwidth]{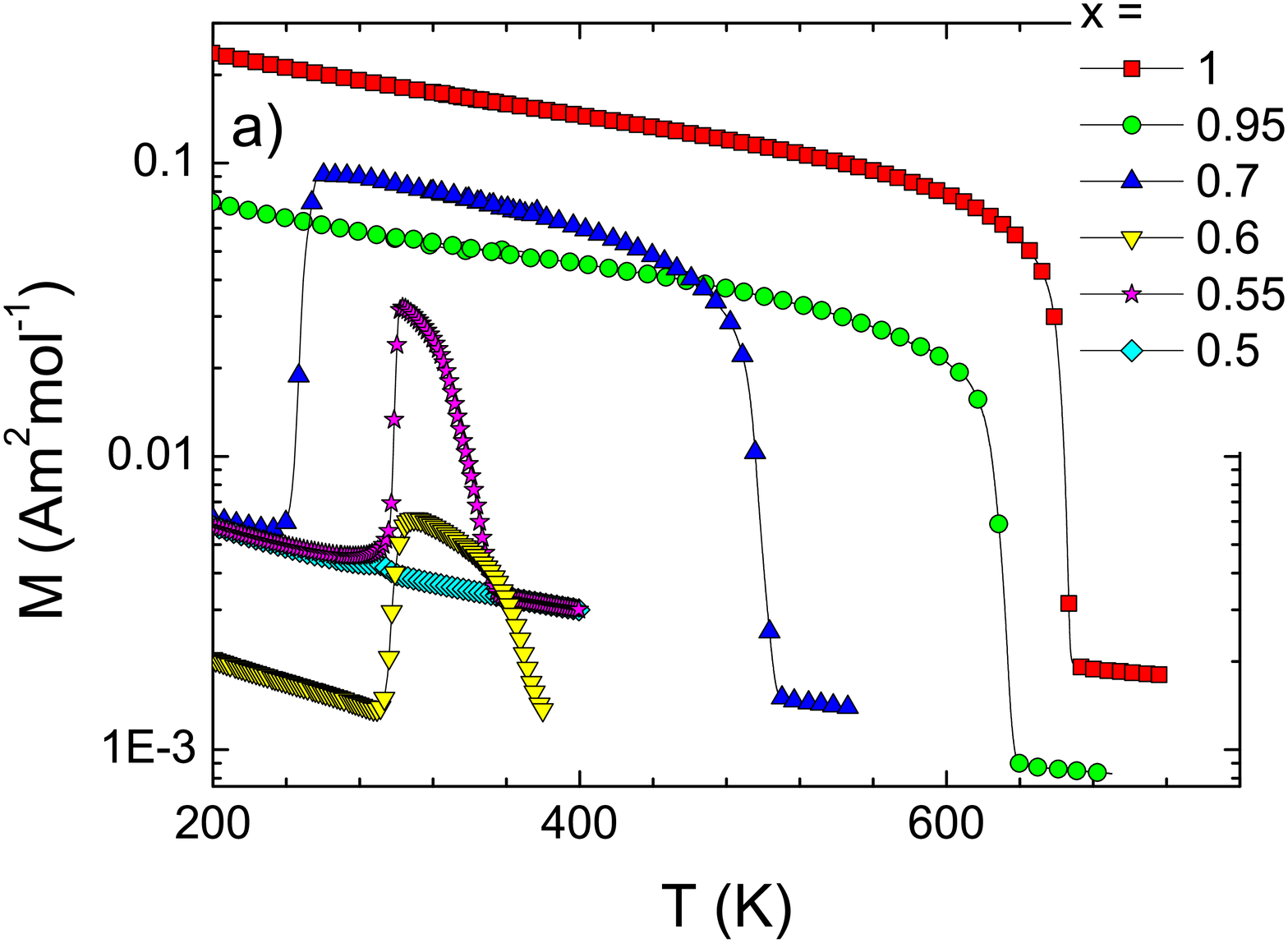}
\end{subfigure} \hspace{0.05\textwidth}
\begin{subfigure}{0.45\textwidth}
\includegraphics[angle=0,width=0.9\textwidth]{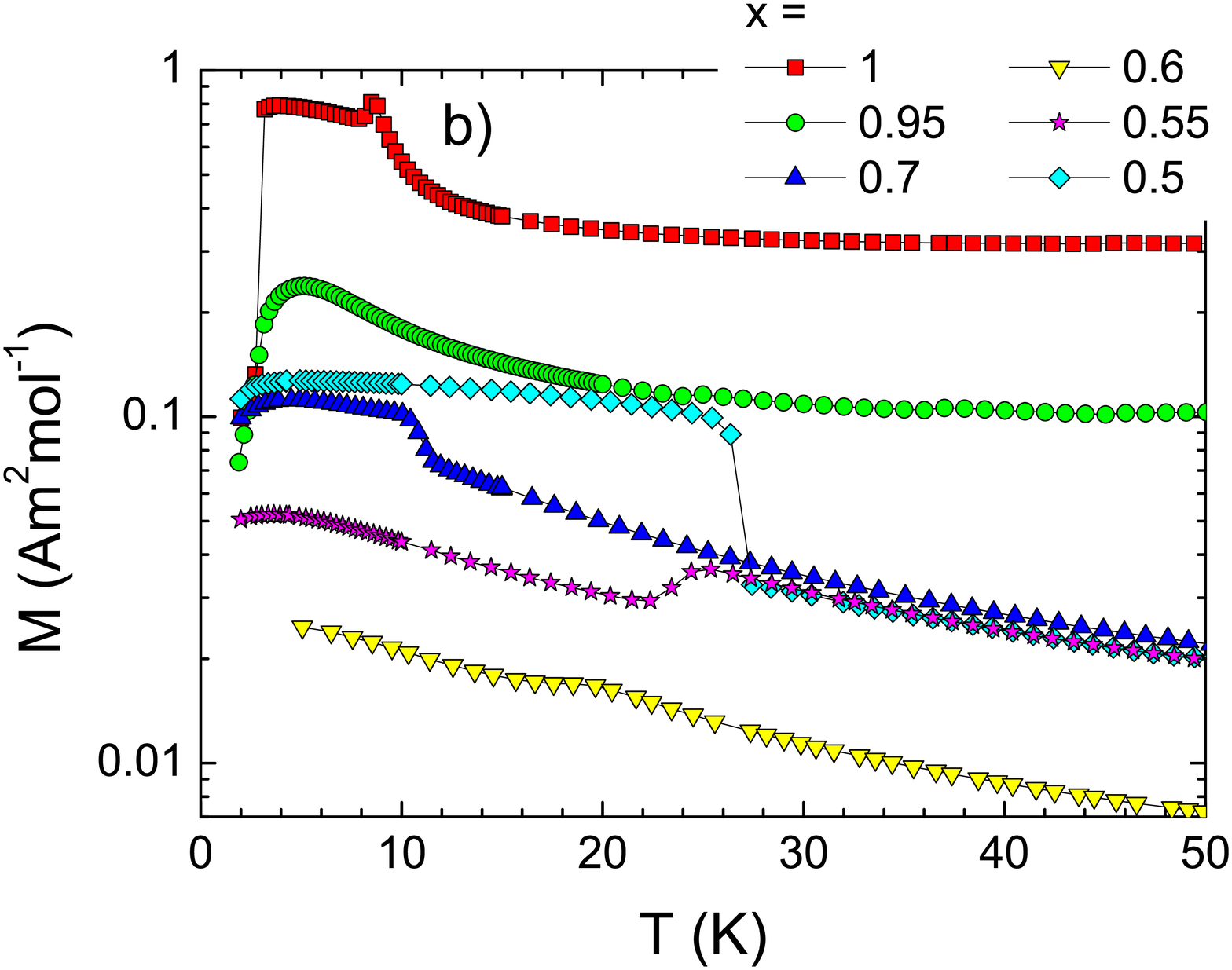}
\end{subfigure}
\caption{The magnetization data measured in applied magnetic field $\mu_0$H = 0.01 T for $x$ = 0.5; 0.55; 0.7; 0.95 and 1 and in applied field $\mu_0$H = XXX T for x = 0.6. a) detail around the ordering temperatures; b) low temperature detail. \label{fig3}}
\pdfcompresslevel=9
\end{center}
\end{figure*}
%
%

%
%
\begin{figure*}[t]
\begin{center}
\begin{subfigure}{0.45\textwidth}
\includegraphics[angle=0,width=0.9\textwidth]{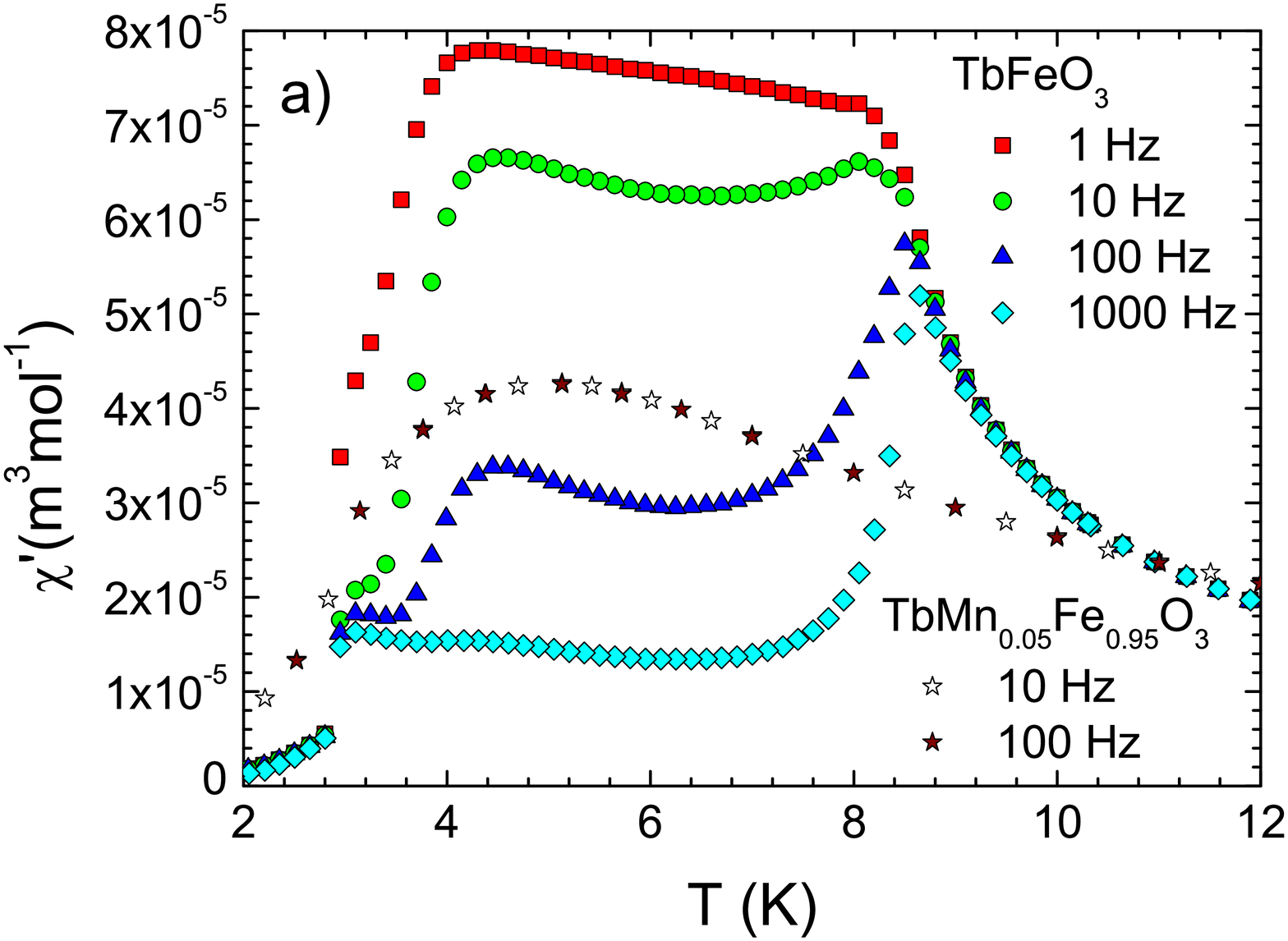}
\end{subfigure} \hspace{0.05\textwidth}
\begin{subfigure}{0.45\textwidth}
\includegraphics[angle=0,width=0.9\textwidth]{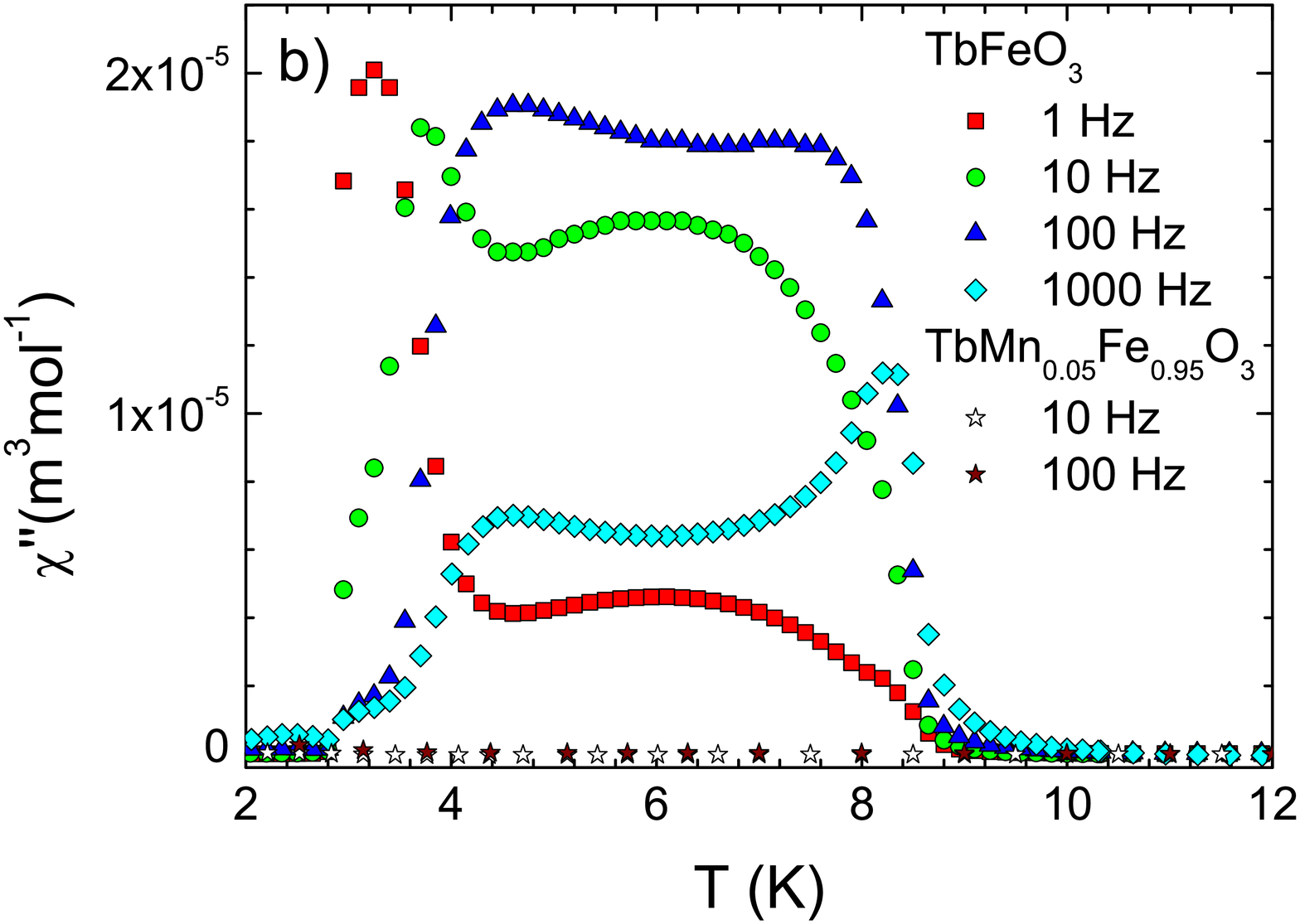}
\end{subfigure}
\caption{The AC susceptibility for TbMn$_{0.05}$Fe$_{0.95}$O$_3$ and TbFeO$_3$ measured with different driving frequencies. a) the real part; b) the imaginary part.  \label{fig4}}
\pdfcompresslevel=9
\end{center}
\end{figure*}
%
%

For concentration $x$ = 0.3 we have observed two anomalies: at 91.4 K and 42 K \cite{mihalik2015}. These anomalies do not fit on general trend which was observed for concentration range $0 \leq x \leq 0.2$. These anomalies occur at too high temperatures to be assigned to ordering of Tb, or Mn sublattice. Since the only remaining magnetic ion is iron, we assign these anomalies to ordering of Fe sublattice. Additional proof that the upper critical temperature is effect of iron sublattice is the magnetization measured for concentration $x$ = 0.4 (Fig. \ref{fig1}). For this concentration we have found critical temperatures 33.4 and 198 K, i.e. paramagnetic to magnetic ordering state critical temperature increases with increasing of iron content.

Temperature dependent magnetization measurements on TbMn$_{1-x}$Fe$_x$O$_3$ for $x \leq 0.5$ revealed that the transition from paramagnetic state to magnetically ordered state increases with the increasing iron content (Fig. \ref{fig3}a). We define the transition temperature as a minimum of $\partial M/\partial T$, which amounts 300 K; 330 K; 366 K; 494 K, 626 K and 661 K for $x$ = 0.5; 0.55; 0.6; 0.7; 0.95 and 1, respectively. Our results correlate with already published data on TbMn$_{0.25}$Fe$_{0.75}$O$_3$ with the magnetic ordering transition at 550 K \cite{kim2011}. The steep decrease of magnetization on cooling, which is connected with order-to-order magnetic phase transition at 300 K, 297 K and 250 K, is typical feature of samples with $x$ = 0.55; 0.6 and 0.7 (Fig. 3\ref{fig3}a). A very similar transition at 180 K was investigated on TbMn$_{0.25}$Fe$_{0.75}$ by Kim et al. \cite{kim2011} using M\"ossbauer spectroscopy and was associated with the spin reorientation effect. In the case of $x$ = 0.5, we have observed at elevated temperatures only the paramagnetic-to magnetically ordered phase transition at 300 K. The single crystal with the same composition and the ordering temperature at 286 K was previously studied by Nhalil et al. \cite{nhalil2015}. The difference in the magnetic phase transition can be probably attributed to slightly different chemical composition because the sample preparation was similar â€“ floating zone method. Nair et al. \cite{nair2016} performed the neutron powder diffraction experiment (NPD) also on $x$ = 0.5 sample and the result from their experiment was the nonzero magnetic contribution to NPD at 295 K and it was found that the magnetic ordering does not take place at exact temperature, but in temperature range 280 -- 303 K. All these findings suggest that the triple point of two magnetic phases with the paramagnetic phase probably exists somewhere in the concentration range $0.5 < x < 0.55$.
 
At low temperatures, we have observed the steep increase of magnetization, in magnetically ordered state for $x$ = 0.5 and $x$ = 0.7 at 27.4 K and 10.9 K; a peak for $x$ = 0.55 and 0.6 at 23.5 K and 18.5 K and two distinct anomalies at 8.3 K and 2.8 K for TbFeO$_3$ (Fig. \ref{fig3}b). The last two anomalies coincide with the anomalies already observed \cite{mihalik2015, bertaut1967} and were explained as Fe spin reorientation at 8.4 K and ordering of Tb sublattice at 3.1 K \cite{bertaut1967}. The anomaly for $x$ = 0.5 was also studied by NPD and it was found that it is order-to-order magnetic phase transition of Mn/Fe (Mn) sublattice at $T_{SR}^{Fe/Mn(Mn)}$ = 26 K \cite{nair2016}.  Depending on ZFC or FC regime the transition can be accompanied by steep decrease or increase of magnetization \cite{ mihalik2015, nair2016} and by narrow peak in heat capacity \cite{nair2016}. Our measurements indicated that the peak in heat capacity is washed out by magnetic field (see Supplementary data), which emphasize the magnetic character of this phase transition. In the case of $x$ = 0.95 we have observed no distinct anomalies in magnetically ordered state neither by magnetization measurements (Fig. \ref{fig3}b) nor by AC susceptibility (Fig. \ref{fig4}). Also, Kim et al. \cite{kim2011} observed no phase transitions for $x$ = 0.75 for $T <$ 100 K. This points out that there exist at least four different magnetic phases at low temperatures in the concentration range $0.5 < x \leq 1$.
 
The AC susceptibility measured for TbFeO$_3$ confirms two anomalies in both, $\chi\prime_{\rm AC}$ and $\chi\prime\prime_{\rm AC}$, at 3.2 and 8.3 K (Fig. \ref{fig4}). These anomalies are frequency-independent up to driving frequencies of 1 kHz and coincide very well with anomalies observed in DC magnetization. We have observed additional bump in both, real and imaginary part around 4 K. This bump shifts to higher temperatures and smears out with increasing frequency. This points out that the magnetic phase in temperature range 3.2 K $< T <$ 8.3 K is not a static one, but has some internal dynamics. Such a dynamic may point out either to the co-existence of several different magnetic phases, or to the displacement of the domain walls in this temperature range. 

%
%
\begin{figure}[t]
\begin{center}
\includegraphics[angle=0,width=0.45\textwidth]{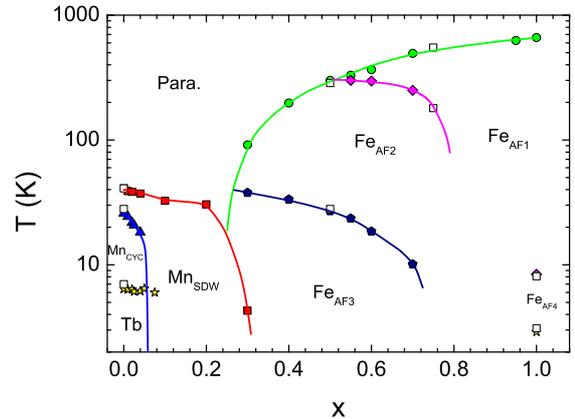}
\caption{The magnetic phase diagram. The full symbols are data presented in this paper; the opened symbols are data from literature \cite{aliouane2008, kim2011, nhalil2015, nair2016, bertaut1967}. \label{fig5}}
\pdfcompresslevel=9
\end{center}
\end{figure}
%
%
All previous findings allowed us to construct the magnetic phase diagram as presented in Fig. \ref{fig5}. We have found two different magnetic phases for Mn sublattices. The ordering temperature for these phases decreases with increasing of Fe concentration. Mn$_{\rm CYC}$ phase (cycloidal magnetic structure, connected with ferroelectric effect \cite{aliouane2008, kimura2005}) vanishes at concentrations lower than $x$ = 0.05, while Mn$_{SDW}$ phase (spin density wave phase \cite{aliouane2008}) persist up to concentration $x \sim 0.3$. At concentration around $x$ = 0.3 we have observed the onset of the ordering of Fe ions into antiferromagnetic Fe$_{\rm AF3}$ phase (G$_x$A$_y$F$_z$ configuration of Fe spins \cite{nhalil2015}) at low temperatures, which coexist with Mn$_{\rm SDW}$ phase at this concentration and simultaneously the formation of antiferromagnetic Fe$_{\rm AF2}$ (A$_x$G$_y$C$_z$ configuration of Fe spins \cite{nhalil2015}) phase at higher temperatures. The further increasing of Fe concentration results to decrease of ordering temperature of Fe$_{\rm AF3}$ phase and simultaneous increase of paramagnetic-to-Fe$_{\rm AF2}$ phase transition. Additional change was observed at concentration around $x$ = 0.5. At this concentration the ordering temperature to Fe$_{\rm AF2}$ phase starts to decrease rapidly with iron content increasing, but additional different antiferromagnetic phase â€“ Fe$_{\rm AF1}$ phase (G$_x$A$_y$F$_z$ type \cite{bertaut1967}) forms at paramagnetic-to-magnetic ordering phase transition. This phase is stable up to $x$ = 1. For parent compound TbFeO$_3$ we have observed also the antiferromagnetic Fe$_{\rm AF4}$ phase (F$_x$C$_y$G$_z$ type \cite{bertaut1967}) at temperatures below 8.2 K. This phase is however unstable and already 5 \% doping of Mn ions into TbFeO$_3$ destabilizes it.  Regarding the magnetic ordering of Tb sublattice, we have found the sings of ordering for parent compounds TbMnO$_3$ and TbFeO$_3$. Already the small substitution of Mn for Fe on iron rich side destroys the magnetic ordering of Tb ions and this ordering does not exist for concentrations $x >$ 0.075 on manganese rich side.
%
%
\section{Conclusions}
Our results have shown that the dominant magnetic ion, which is responsible for the magnetic ordering in the concentration range $0 \leq x < 0.3$ is Mn$^{3+}$, while iron ions do not create independent magnetic structure. The only effect of iron ions is the weakening of the Mn--O--Mn superexchange interaction by replacing some Mn$^{3+}$ ions with Fe$^{3+}$ ions. This leads to decreasing of the critical temperatures for both, Mn$_{\rm CYC}$ and Mn$_{\rm SDW}$ phase. On the other hand, the magnetism in the concentration range $0.3 < x \leq 1$ is driven by the Fe$^{3+}$ ions and we have found four different magnetic phases, but only three different magnetic structures. Pure TbFeO$_3$ undergoes spin reorientation order-to-order magnetic transition and thence there exist at least two different magnetic structures with very similar total energies in this compound. Four magnetic phases can be explained by decreasing of total energy of some magnetic structure caused by manganese ion. That is why we conclude that the manganese ions are responsible for stabilization of four magnetic phases of Fe sublattice in TbMn$_{1-x}$Fe$_x$O$_3$ compound.






\section*{Acknowledgment}
This work was supported by projects VEGA 2/0132/16, SK-PT-2015-0030, ERDF EU under the contract No ITMS26220120047 and Slovenian Re\-search Agency (ARRS), Program Number P2-0348.

\section*{References}

\bibliography{mybibfile}

\end{document}